\documentclass[showpacs,preprintnumbers,amsmath,amssymb,aps,pre,onecolumn]{revtex4}
\usepackage{graphicx}% Include figure files
\usepackage{dcolumn}
\usepackage{bm}
\usepackage{epsfig}

\begin{document}

\title{Time series irreversibility: a visibility graph approach}

\author{Lucas Lacasa$^{1*}$, Angel Nu\~{n}ez$^1$, \'Edgar Rold\'an$^2$, Juan M.R. Parrondo$^2$, and Bartolo Luque$^1$}
\email{lucas.lacasa@upm.es}
\address{$^1$Departamento de Matem\'{a}tica Aplicada y Estad\'{i}stica,
ETSI Aeron\'{a}uticos,  Universidad Polit\'{e}cnica de Madrid, Spain\\
$^2$Departamento de F\'{i}sica At\'{o}mica, Molecular y Nuclear and
\emph{GISC}, Universidad Complutense de Madrid, Spain}

\date{\today}
% It is always \today, today, % but any date may be
% explicitly specified


\begin{abstract}
We propose a method to measure real-valued time series
irreversibility which combines two different tools: the horizontal
visibility algorithm and the Kullback-Leibler divergence. This
method maps a time series to a directed network according to a
geometric criterion. The degree of irreversibility of the series is
then estimated by the Kullback-Leibler divergence (i.e. the
distinguishability) between the \emph{in} and \emph{out} degree
distributions of the associated graph. The method is computationally
efficient, does not require any \emph{ad hoc} symbolization process,
and naturally takes into account multiple scales. We find that the
method correctly distinguishes between reversible and irreversible
stationary time series, including analytical and numerical studies
of its performance for: (i) reversible stochastic processes
(uncorrelated and Gaussian linearly correlated), (ii) irreversible
stochastic processes (a discrete flashing ratchet in an asymmetric
potential), (iii) reversible (conservative) and irreversible
(dissipative) chaotic maps, and (iv) dissipative chaotic maps in the
presence of noise. Two alternative graph functionals, the degree and
the degree-degree distributions, can be used as the Kullback-Leibler
divergence argument. The former is simpler and more intuitive and
can be used as a benchmark, but in the case of an irreversible
process with null net current,  the degree-degree distribution has
to be considered to identifiy the irreversible nature of the series.
\end{abstract}

\maketitle %%%

\section{Introduction}
A stationary process $X(t)$ is said to be statistically time
reversible (hereafter time reversible) if for every $N$, the
series $\{X(t_1), \cdots , X(t_N)\}$ and $\{X(t_N), \cdots ,
X(t_1)\}$ have the same joint probability distributions
\cite{weiss}. Roughly, this means that a reversible time series
and its time reversed are, statistically speaking, equally
probable. Reversible processes include the family of Gaussian
linear processes (as well as Fourier-transform surrogates and
nonlinear static transformations of them), and are associated with
processes at thermal equilibrium in statistical physics.
Conversely, time series irreversibility is indicative of the
presence of nonlinearities in the underlying dynamics, including
non-Gaussian stochastic processes and dissipative chaos, and are
 associated with systems driven out-of-equilibrium in the
realm of thermodynamics \cite{Parrondo0, Parrondo1}. Time series
irreversibility is an important topic in basic and applied
science. From a physical perspective, and based on the relation
between statistical reversibility and physical dissipation
\cite{Parrondo0, Parrondo1}, recent work uses the concept of time
series irreversibility to derive information about the entropy
production of the physical mechanism generating the series, even
if one ignores any detail of such mechanism \cite{Parrondo2,
Parrondo3}. In a more applied context, it has been suggested that
irreversibility in complex physiological series decreases with
aging or pathology, being maximal in young and healthy subjects
\cite{heartbeatPRL,costa1, multiscale}, rendering this feature
important for noninvasive diagnosis. As complex signals pervade
natural and social sciences, the topic of time series
reversibility is indeed relevant for scientists aiming to
understand and model the dynamics behind complex signals.

The definition of time series reversibility is formal and therefore
there is not an \emph{a priori} optimal algorithm to quantify it in
practice. Recently,  several methods to measure time irreversibility
have been proposed  \cite{kennel2, kennel, diks, gaspard, costa1,
cammarota, heartbeatPRL, andrieux, wang}. The majority of them
perform a time series symbolization, typically making an empirical
partition of the data range \cite{kennel2} (note that such a
transformation does not alter the reversible character of the output
series \cite{kennel}) and subsequently analyze the symbolized
series, through statistical comparison of symbol strings occurrence
in the forward and backwards series or using
 compression algorithms \cite{kennel,
Parrondo3,cover}. The first step requires an extra amount of
\emph{ad hoc} information (such as range partitioning or size of the
symbol alphabet) and therefore the output of these methods
eventually depend on these extra parameters. A second issue is that
since typical symbolization is local, the presence of multiple
scales (a signature of complex signals) could be swept away by this
coarse-graining: in this sense multi-scale algorithms have been
proposed recently \cite{costa1, multiscale}.

Motivated by these facts, here we explore the usefulness of the
horizontal visibility algorithm  in such context. This is a time
series analysis method which was  proposed recently \cite{HVA}. It
makes use of graph theoretical concepts, and it is based on the
mapping of a time series to a graph and the subsequent analysis of
the associated graph properties \cite{pnas,HVA,epl,toral}. Here we
propose a \emph{time directed} version of the horizontal visibility
algorithm, and we show that it is a simple and well defined tool for
measuring time series irreversibility. More precisely, we show that
the Kullback-Leibler divergence \cite{cover} between the \emph{out}
and \emph{in} degree distributions, $D[P_{\rm out}(k)||P_{\rm
in}(k)]$, is a simple measure of the irreversibility of real-valued
stationary stochastic series. Analytical and numerical results
support our claims, and the presentation is as follows: The method
is introduced in section II. In section III we analyze reversible
time series generated from linear stochastic processes, which yield
$D[P_{\rm out}(k)||P_{\rm in}(k)]=0$. As a further validation, in
section IV we report the results obtained for irreversible series.
We first analyze a thermodynamic system (a discrete flashing
ratchet) which shows time irreversibility when driven out of
equilibrium. Its amount of irreversibility can be increased
continuously tuning the value of a parameter of the system, and we
find that the method can, not only distinguish, but also quantify
the degree of irreversibility. We also study the effect of applying
a stalling force in the opposite direction of the net current of
particles in the ratchet. In this case the benchmark measure fails
predicting reversibility whereas a generalized measure based on
degree-degree distributions $D[P_{\rm out}(k,k')||P_{\rm in}(k,k')]$
goes beyond the phenomenon associated to physical currents and still
detects irreversibility. We extend this analysis to chaotic signals,
where our method distinguishes between dissipative and conservative
chaos, and we analyze chaotic signals polluted with noise. Finally,
a discussion is presented in section V.

\section{The method}
\subsection{The horizontal visibility graph}

The family of visibility algorithms is a collection of methods
that map series to networks according to specific geometric
criteria \cite{pnas, HVA}. The general purpose of such methods is
to accurately map the information stored in a time series into an
alternative mathematical structure, so that the powerful tools of
graph theory may eventually be employed to characterize time
series from a different viewpoint, bridging the gap between
nonlinear time series analysis, dynamical systems, and graph
theory \cite{epl,plos,elsner,turbulence,finance}.

We focus here on a specific subclass called horizontal visibility
algorithm, firstly proposed in \cite{HVA}, and defined as follows:
Let $\{x_t\}_{t=1,. . .,N}$ be a real-valued time series of $N$
data. The algorithm assigns each datum of the series to a node in
the {\sl horizontal visibility graph} (HVg). Then, two nodes $i$ and
$j$ in the graph are connected if one can draw a \emph{horizontal}
line in the time series joining $x_i$ and $x_j$ that does not
intersect any intermediate data height (see figure \ref{imagen}).
Hence, $i$ and $j$ are two connected nodes if the following
geometrical criterion is fulfilled within the time series:
\begin{equation}
x_i,x_j > x_n, \ \forall \ n \ \left | \ i < n < j\right.  \label{criterio}
\end{equation}
Some results regarding the characterization of stochastic and
chaotic series through this method have been put forward recently
\cite{HVA,toral}, and the first steps for a mathematically sound
characterization of horizontal visibility graphs have been
established \cite{simone}. Interestingly, a very recent work
suggests that the method can be used in practice to characterize not
only time series but generic nonlinear discrete dynamical systems,
sharing similarities with the theory of symbolic dynamics
\cite{plos}.

\subsection{Directed HVg}

So far in the literature the family of visibility graphs are
undirected, as visibility did not have a predefined temporal arrow.
However, as already suggested in the seminal paper \cite{pnas}, such
a directionality can be made explicit by making use of directed
networks or digraphs \cite{redes}. We address such directed version,
defining a \emph{Directed} Horizontal Visibility graph (DHVg) as a
HVg, where the degree $k(t)$ of the node $t$ is now splitted in an
\emph{ingoing} degree $k_{\rm in}(t)$, and an \emph{outgoing}
degree, such that $k(t)=k_{\rm in}(t)+k_{\rm out}(t)$. The ingoing
degree $k(t)$ is defined as the number of links of node $t$ with
other \emph{past} nodes associated with data in the series  (that
is, nodes with $t'<t$). Conversely, the outgoing degree $k_{\rm
out}(t)$, is defined as the number of links with \emph{future}
nodes.

\begin{figure}[h]
\centering
\includegraphics[width=0.7\textwidth]{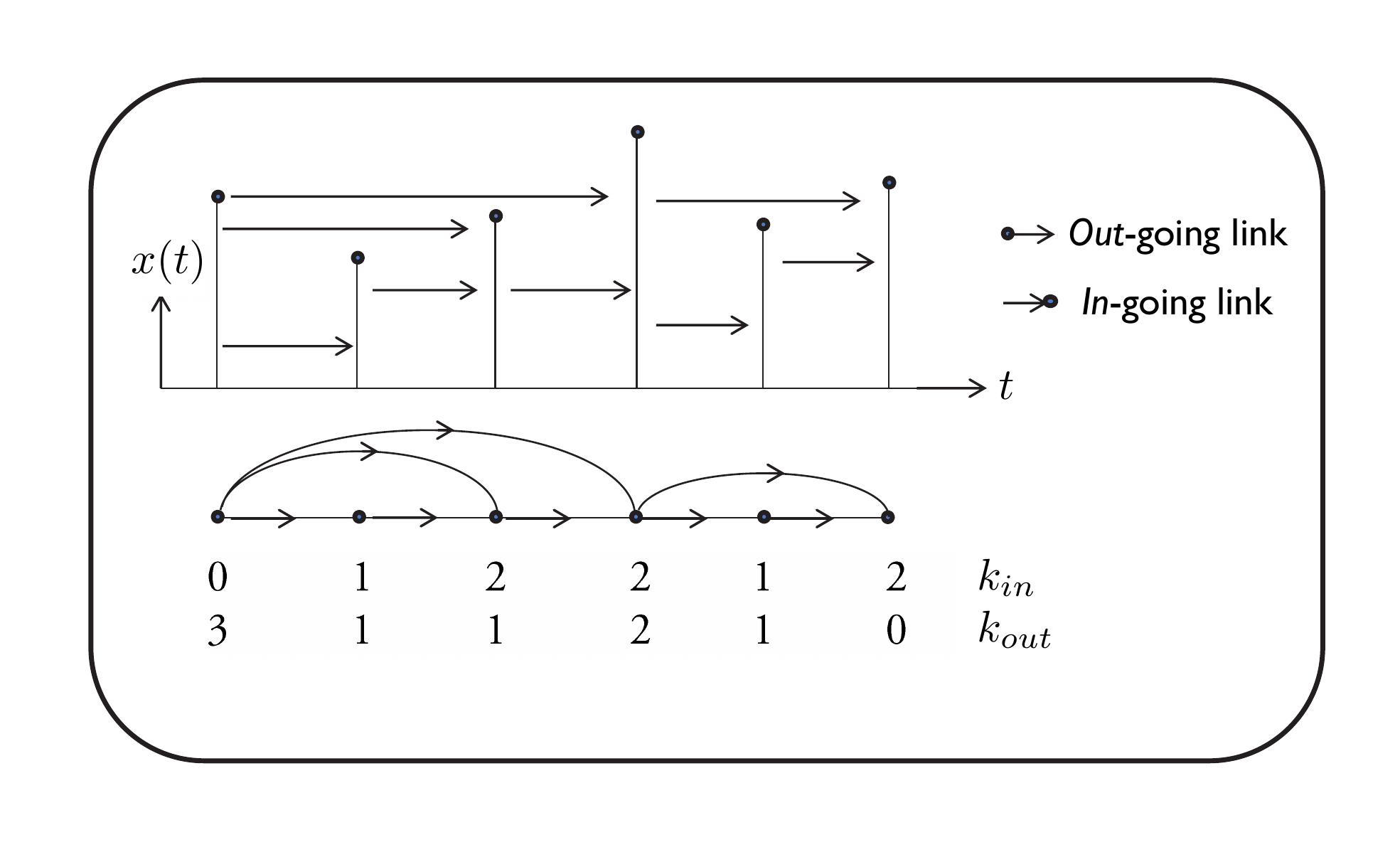}
\caption{Graphical illustration of the method. In the top we plot
a sample time series $\{x(t)\}$. Each datum in the series is
mapped to a node in the graph. Arrows, describing allowed directed
visibility, link nodes. The associated directed horizontal visibility graph
is plotted below. In this graph, each node has an ingoing degree
$k_{\rm in}$, which accounts for the number of links with $past$
nodes, and an outgoing degree $k_{\rm out}$, which in turn accounts
for the number of links with $future$ nodes. The asymmetry of the
resulting graph can be captured in a first approximation through
the invariance of the outgoing (or ingoing) degree series under time reversal.}
\label{imagen}
\end{figure}

For a graphical illustration of the method, see figure \ref{imagen}.
The degree distribution of a graph describes the probability of an
arbitrary node to have degree $k$ (i.e. $k$ links) \cite{redes}. We
define the \emph{in} and \emph{out} (or ingoing and outgoing) degree
distributions of a DHVg as the probability distributions of $k_{\rm
out}$ and $k_{\rm in}$ of the graph which we call $P_{\rm
out}(k)\equiv P(k_{\rm out}=k)$ and $P_{\rm in}(k)\equiv P(k_{\rm
in}=k)$, respectively.

\subsection{Quantifying irreversibility: DHVg and Kullback-Leibler divergence}

 The main conjecture of this work is that the information stored in the $in$ and $out$
 distributions take into account the
amount of time irreversibility of the associated series. More
precisely, we claim that this can be measured, in a first
approximation, as the distance (in a distributional sense) between
the $in$ and $out$ degree distributions ($P_{\rm in}(k)$ and $P_{\rm
out}(k)$). If needed, higher order measures can be used, such as the
corresponding distance between the $in$ and $out$ degree-degree
distributions ($P_{\rm in}(k,k')$ and $P_{\rm out}(k,k')$). These
are defined as the $in$ and $out$ joint degree distributions of a
node and its first neighbors \cite{redes}, describing the
probability of an arbitrary node whose neighbor has degree $k'$ to
have degree $k$.

We make use of the Kullback-Leibler divergence \cite{cover} as the
distance between the \emph{in} and \emph{out} degree distributions.
Relative entropy or Kullback-Leibler divergence (KLD) is introduced
in information theory as a measure of distinguishability between two
probability distributions.
 Given a random variable $x$ and two probability distributions $p(x)$ and $q(x)$, KLD between
$p$ and $q$ is defined as follows:

\begin{equation}
D(p||q)\equiv\sum_{x\in\mathcal{X}} p(x)\log\frac{p(x)}{q(x)},
\end{equation}
 which vanishes if and only if both probability distributions are equal $p=q$
and it is bigger than zero otherwise. Unlike other measures used to
estimate time irreversibility \cite{kennel2, kennel, cammarota,
heartbeatPRL}, the KLD is statistically significant, as it is proved
by the Chernoff-Stein lemma: The probability of incorrectly guessing
(via hypothesis testing) that a sequence of $n$ data is distributed
as $p$  when the true distribution is $q$ tends to $e^{-nD(p||q)}$
when $n\to\infty$. The KLD is then related to the probability to
fail when doing an hypothesis test, or equivalently, it is a measure
of distinguishability: the more distinguishable are $p$ and $q$ with
respect to each other, the larger is $D(p||q)$.

In statistical mechanics, the KLD can be used to measure the time
irreversibility of data produced by nonequilibrium processes
 but also to estimate the average entropy production of the physical process that
generated the data \cite{Parrondo0,Parrondo2}.   Irreversibility can
be assessed  by the KLD between  probability distributions
associated to observables in the process and in its time reversal.
These measure gives lower bounds to the entropy production, whose
accuracy increases as the observables contain a more detailed
description of the system. The measure that we present in this work
has this limitation: it takes the information from the degree, which
is  a partial description of the process. Consequently, our
technique does not give a tight bound to the entropy production.

Nevertheless, as we will show in several examples, the information
of the outgoing degree distribution $k_{\rm out}$ is sufficient to
distinguish between reversible and irreversible stochastic
stationary series which are real-valued but discrete in time $\{ x_t
\}_{t=1,...,N}$. We compare the outgoing degree distribution in the
actual (forward) series $P_{k_{\rm out}} (k| \{ x(t) \}_{t=1,...,N} )=
P_{\rm out}(k)$ with the corresponding probability in the
time-reversed (or backward) time series, which is equal to the
probability distribution of the ingoing degree in the actual process
$P_{k_{\rm out}}(k|\{x(t)\}_{t=N,...,1}) = P_{\rm in}(k)$. The KLD
between these two distributions is
\begin{equation}
D[P_{\rm out}(k)||P_{\rm in}(k)]=\sum_k
P_{\rm out}(k)\log\frac{P_{\rm out}(k)}{P_{\rm in}(k)}.\label{dkl}
\end{equation}
This measure vanishes if and only if the outgoing and ingoing degree
probability distributions of a time series are identical, $P_{\rm
out}(k)=P_{\rm in}(k)$, and it is positive otherwise. We will apply
it to several examples as a measure of irreversibility.

Notice that previous methods to estimate time series irreversibility
generally proceed by first making a (somewhat \emph{ad hoc}) local
symbolization of the series, coarse-graining each of the series data
into a symbol (typically, an integer) from an ordered set. Then,
they subsequently perform a statistical analysis of word occurrences
(where a word of length $n$ is simply a concatenation of $n$
symbols) from the forward and backwards symbolized series
\cite{andrieux,wang}. Time series irreversibility is therefore
linked to the difference between the word statistics of the forward
and backwards symbolized series. The method presented here can also
be considered as a symbolization if we restrict ourselves to the
information stored in the series $\{k_{\rm out}(t)\}_{t=1,...,N}$ and
$\{k_{\rm in}(t)\}_{t=1,...,N}$. However, at odds with other methods,
here the symbolization process (i) lacks \emph{ad hoc} parameters
(such as number of symbols in the set or partition definition), and
(ii) it takes into account \emph{global} information: each
coarse-graining $x_t\rightarrow (k_{\rm in}(t),k_{\rm out}(t))$ is
performed using information from the whole series, according to the
mapping criterion \eqref{criterio}. Hence, this symbolization
naturally takes into account multiple scales, which is desirable if
we want to tackle complex signals \cite{costa1,multiscale}.

\section{Reversibility}
\subsection{Uncorrelated stochastic series}
\begin{figure}[h] \centering
\includegraphics[width=0.7\textwidth]{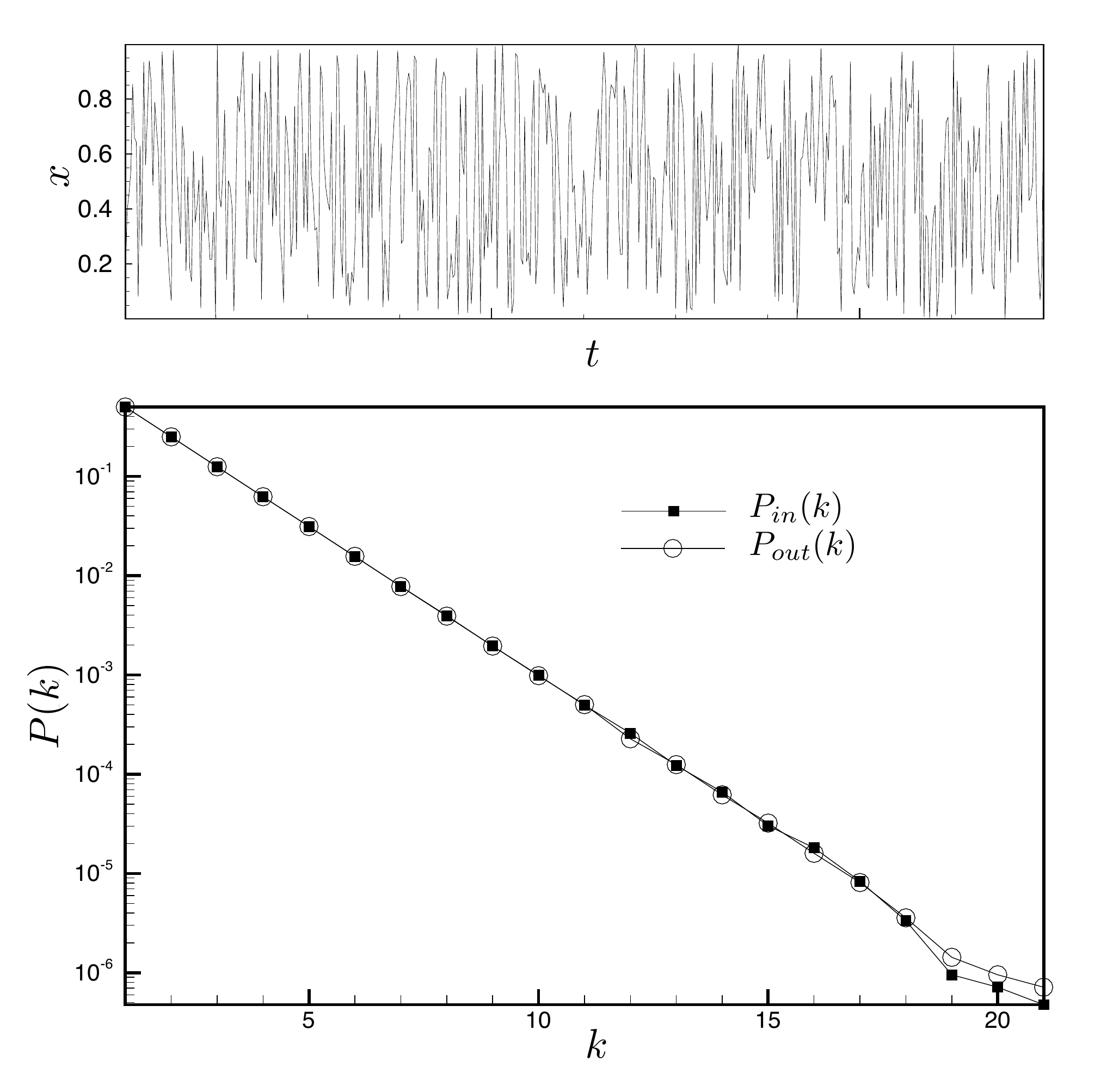}
\caption{\emph{Top}: A sample uncorrelated random time series
extracted from a uniform distribution $U[0,1]$. \emph{Bottom}: The
\emph{in} and \emph{out} degree distributions of the DHVg
associated to the random series. The process is reversible and the
graph degree distributions are, besides finite size effects,
equivalent. The deviation is measured through their KLD (see table
\ref{table1}).} \label{random}
\end{figure}

For illustrative purposes, in figure \ref{random} we have plotted
the \emph{in} and \emph{out} degree distributions of the visibility
graph associated to an uncorrelated random series $\{ x_t
\}_{t=1,\dots,N}$ of size $N=10^6$ : the distributions cannot be
distinguished and KLD vanishes (the numerical value of KLD is shown
in table \ref{table1}) which is indicative of a reversible series.
In what follows we provide an exact derivation of the associated
outgoing and ingoing degree distributions associated to this
specific process, showing that they are indeed identical in the
limit of infinite size series.

  \textsf{Theorem 1.} Let
$\{x_{t}\}_{t=-\infty,\dots,\infty}$ be a bi-infinite sequence of
independent and identically distributed random variables extracted
from a continuous probability density $f(x)$. Then, both the
\emph{in} and \emph{out} degree distributions of its associated
directed horizontal visibility graph are
\begin{equation}
P_{\rm in}(k)=P_{\rm out}(k)=\bigg(\frac{1}{2}\bigg)^{k},\ k=1,2,3,...
\label{pk}
\end{equation}

  \textsf{Proof (out-distribution).}  Let $x$ be an
arbitrary datum of the aforementioned series. The probability that
the horizontal visibility of $x$ is interrupted by a datum $x_{r}$
on its right is independent of $f(x)$,

\begin{equation}
\Phi _{1}=\int_{-\infty }^{\infty }\int_{x}^{\infty
}f(x)f(x_{r})dx_{r}dx\nonumber\\
=\int_{-\infty }^{\infty }f(x)[1-F(x)]dx=\frac{1}{2}, \label{phi}
\end{equation}%
where $F(x)=\int_{-\infty}^x f(x')dx'$.

The probability $P(k)$ of
the datum $x$ being capable of exactly seeing $k$ data may be
expressed as
\begin{equation}
P(k)=Q(k)\Phi _{1}=\frac{1}{2}Q(k), \label{recurrence}
\end{equation}%
where $Q(k)$ is the probability of $x$ seeing at least $k$ data.
$Q(k)$ may be recurrently calculated via
\begin{equation}
Q(k)=Q(k-1)(1-\Phi _{1})=\frac{1}{2}Q(k-1),
\end{equation}%
from which, with $Q(1)=1$, the following expression is obtained
\begin{equation}
Q(k)=\bigg(\frac{1}{2}\bigg)^{k-1},
\end{equation}%
which together with equation (\ref{recurrence}) concludes the proof.
An analogous derivation holds for the \emph{in} case.

Note that this result is independent of the underlying probability
density $f(x)$: it holds not only for Gaussian or uniformly
distributed random series, but for any series of independent and
identically distributed (i.i.d.) random variables extracted from a
continuous distribution $f(x)$. A trivial corollary of this theorem
is that the KLD between the \emph{in} and \emph{out} degree
distributions associated to a random uncorrelated process tends
asymptotically to zero with the series size, which correctly
suggests that the series is time reversible.

\subsection{Correlated stochastic series}

\begin{figure}[h]
\centering
\includegraphics[width=0.6\textwidth]{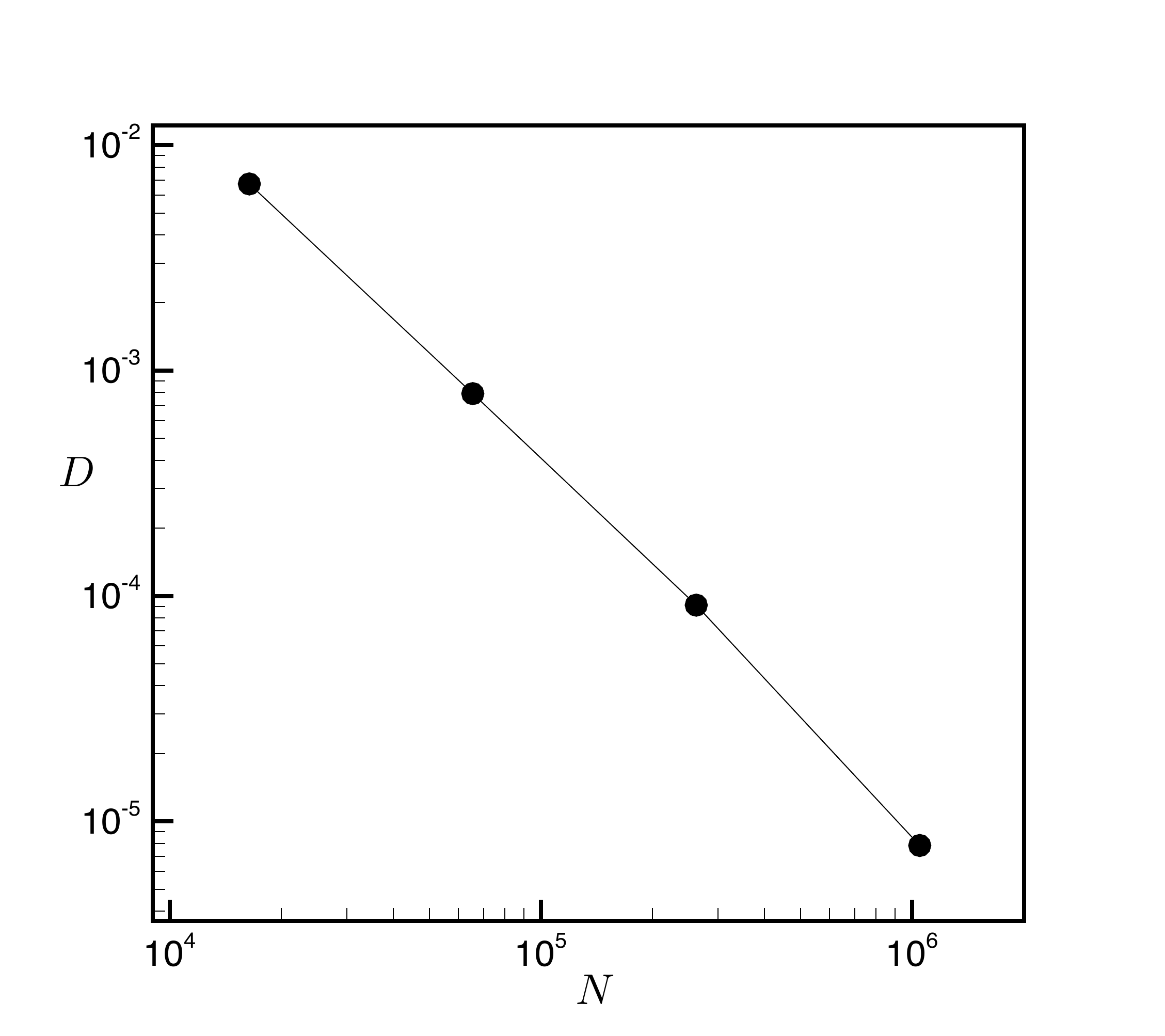}
\caption{Log-log plot of $D[P_{\rm out}(k)||P_{\rm in}(k)]$ of the
graph associated to an Ornstein-Uhlenbeck process  as a function
of the series size $N$ (dots are the result of an ensemble average
over  several realizations). Note that KLD
 decreases with series size and  tends to zero asymptotically.}
\label{finitesize}
\end{figure}
In the last section we considered uncorrelated stochastic series
which are our first example of a reversible series with $D[P_{\rm
out}(k)||P_{\rm in}(k)]=0$.
 As a further validation, here we focus on linearly
correlated stochastic processes as additional examples of
reversible dynamics \cite{weiss}. We use the \emph{minimal substraction procedure} \cite{toral}
 to generate such correlated series. This method is a modification of
the standard Fourier filtering method, which consists in filtering a series of
uncorrelated random numbers in Fourier space.
% The classical way to
%generate such correlated series is to filter a series of
%uncorrelated random numbers in Fourier space (the so-called
%Fourier filtering method), but we make
%use of a modification of this method called the
%\emph{minimal substraction procedure} \cite{toral}.
%Such method
%considers the issue that not every function $C(t)$ can be
%considered to be the correlation function of a Gaussian field,
%since some mathematical requirements need to be fulfilled, namely
%that the quadratic form $\sum_{ij} x_iC(|i-j|)x_j$ be positive
%definite, and accordingly applied a regularization procedure on
%$C(t)$, enabling the generation of correlated Gaussian random
%numbers $x_i$ of zero mean and correlation function $\langle
%x_ix_j\rangle=C(|i-j|)$.
We study time series whose correlation is exponentially decaying $C(t)\sim
\exp(-t/\tau)$ (akin to an Ornstein-Uhlenbeck process) and power
law decaying $C(t)\sim t^{-\gamma}$. % and we
%calculate our irreversibility measure on their associated
%visibility graphs. In table \ref{table1} %(concretely, for the series with
%exponentially decaying correlations we set $\tau=1$, whereas in
%the one with power-law decaying correlations we set $\gamma=2$).
In table \ref{table1}  we show that the KLD of these series are all very close to zero, %in good agreement with
%the reversible character of the respective series.
 and its deviation from zero is originated by finite size effects, as it is shown in figure \ref{finitesize}.

\section{Irreversibility}

\subsection{Discrete flashing ratchet}

\begin{figure}[h]
\centering
\includegraphics[width=0.6\textwidth]{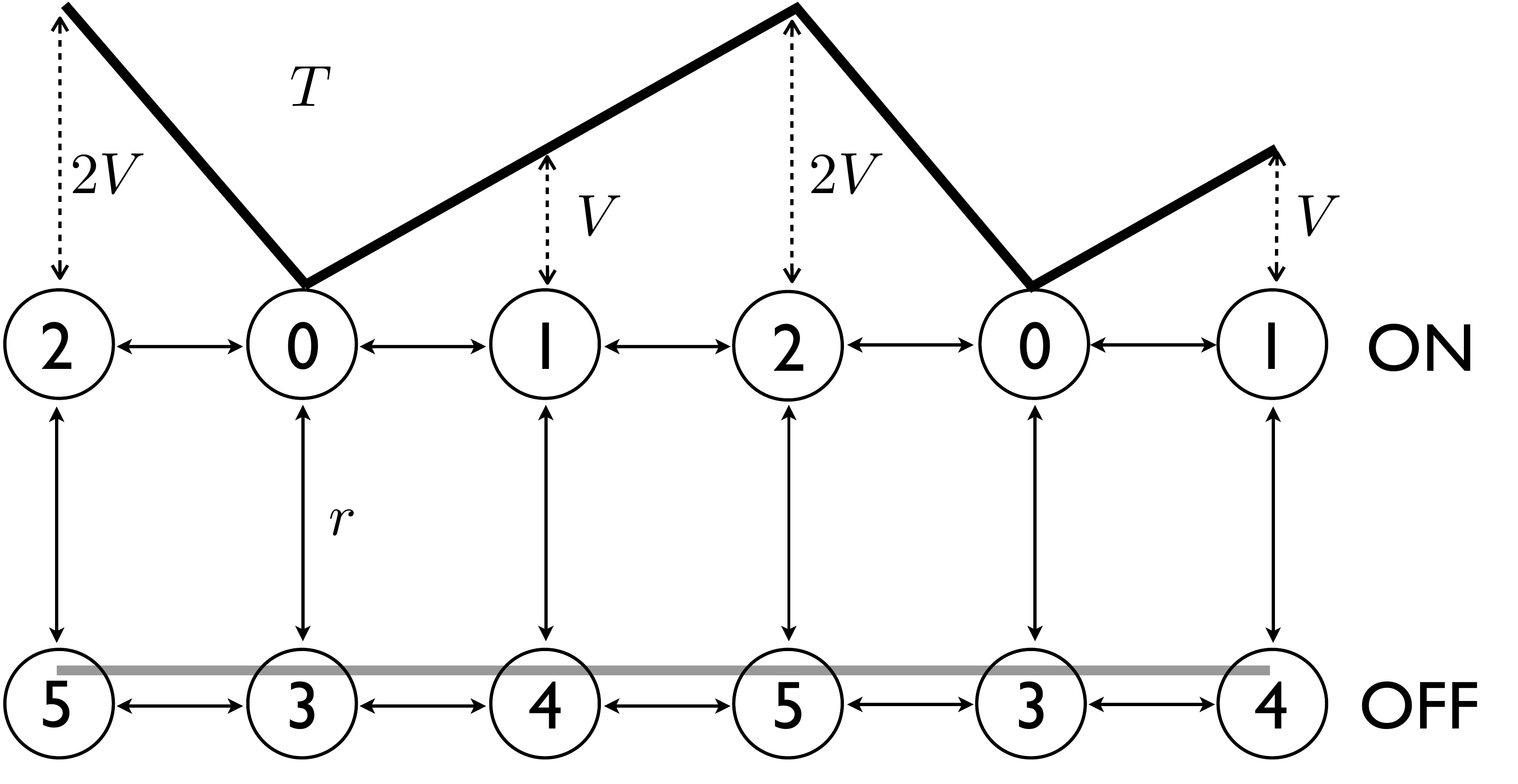}
\caption{Discrete flashing ratchet scheme. Particles are at
temperature $T$ moving in a periodic linear asymmetric potential
of height $2V$. The potential is switched on and off at a constant
rate $r$, which originates a net current of particles to the left.
If the potential is ON, the state of the potential is
represented by its position $x=\{0, 1, 2\}$, and if it is OFF by
$x+3=\{3,4,5 \}$.} \label{drfig}
\end{figure}

\begin{figure}[h]
\centering
\includegraphics[width=0.6\textwidth]{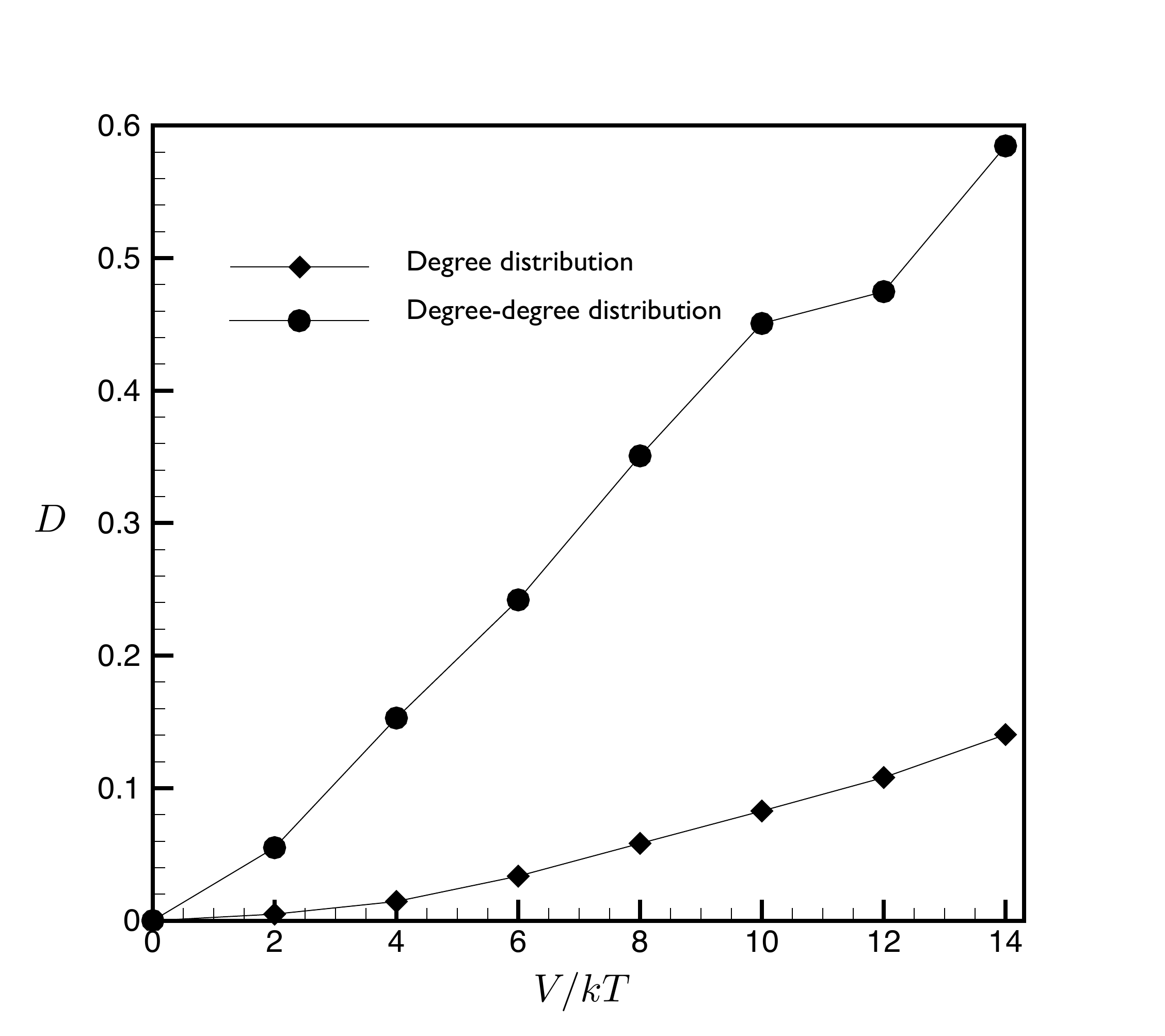}
\caption{$D[P_{\rm out}(k)||P_{\rm in}(k)]$ and
$D[P_{\rm out}(k,k')||P_{\rm in}(k,k')]$ for a discrete flashing ratchet
($r=1$) as a function of $V/kT$. For each value
of $V$ we generate a stationary time series of
$N=10^6$ steps described with full information (position and state of the potential).
The system is in equilibrium for $V=0$, and it is driven out of equilibrium for $V>0$.} \label{discreta}
\end{figure}

 We now study a thermodynamic system which can be smoothly driven
out of equilibrium by modifying the value of a physical parameter.
We make use of the time series generated by
  a discrete flashing ratchet model introduced in \cite{Parrondo2}. The ratchet
consists of a particle moving in a one dimensional lattice. The
particle is at temperature $T$ and moves in a periodic asymmetric
potential of height $2V$, which is switched on and off at a rate $r$
(see Figure \ref{drfig} for details). The switching rate is
independent of the position of the particle, breaking detailed
balance \cite{Parrondo2,Parrondo3}. Hence, switching the potential
drives the system out of equilibrium resulting in a directed motion
or net current of particles. When using full information of the
process, trajectories of the system are described
 by two variables: the position of the particle $x=\{0,1,2\}$ and the state of the
potential, $y=\{\textrm{ON},\textrm{OFF}\}$. The time series are
constructed from $x$ and $y$ variables as follows: $(x,y)=x\;$ if
$y=\textrm{ON}$ and $(x,y)=x+3\;$ if $y=\textrm{OFF}$.

The dynamics of the system is described by  a six-state Markov chain
with transition probabilities $p_{i\to j}=\Gamma_{i\to j}/\sum_{j}
\Gamma_{i\to j}$, where $\Gamma_{i\to j}$ is the transition rate
from $i$ to $j$ and the sum $\sum_{j}$ runs over the accessible
states from $i$ (see figure \ref{drfig}). All transition rates
satisfy the detailed balance condition with respect to the thermal
bath at temperature $T$, except the switches between ON and OFF.
When the potential is on, $i,j=\{0,1,2\}$ and $\Gamma_{i\to j}=\exp
[- (V_j -V_i)/kT]$. When it is off, $i,j=\{3,4,5\}$ and
$\Gamma_{i\to j}=1$. On the other hand, switches are implemented
with rates that do not depend on the position of the particle and
therefore do not satisfy detail balance condition \cite{Parrondo3}:
$\Gamma_{i\to i+3}=\Gamma_{i+3\to i}=r$, for $i=\{0,1,2\}$.

%%Dynamics are implemented with a Metropolis algorithm
%%where the transition probability goes like $e^{-\beta \Delta V}$,
%%where $\beta=1/kT$ and $\Delta V$ is the potential \verb"difference"
%%between the states.

In Figure \ref{discreta} we depict the values of $D[P_{\rm
out}(k)||P_{\rm in}(k)]$ and $D[P_{\rm out}(k,k')||P_{\rm
in}(k,k')]$ as a function of $V/kT$,
 for $6-$state time series of $2^{19}$ data. Note
that for $V=0$ detailed balance condition is satisfied, the
system is in equilibrium and trajectories are statistically
reversible. In this case both KLD using degree distributions and degree-degree
distributions vanish. On the other hand, if $V$ is increased,
 the system is driven out of equilibrium, what
introduces a net statistical irreversibility which increases with
$V$ \cite{Parrondo2}. The amount of irreversibility estimated with
KLD increases with $V$ for both measures, therefore the results
produced by the method are qualitatively correct. Interestingly
enough, the tendency holds even for high values of the potential,
where the statistics are poor and the KLD of sequences of symbols
usually fail when estimating irreversibility \cite{Parrondo2}.
However the values of the KLD that we find are far below the KLD per
step between the forward and backward trajectories, which is equal
to the dissipation as reported in \cite{Parrondo2}. The degree
distributions capture the irreversibility of the original series but
it is difficult to establish a quantitative relationship between
\eqref{dkl} and the KLD between trajectories.

On the other hand, the measure based on the degree-degree
distribution $D[P_{\rm out}(k,k')||P_{\rm in}(k,k')]$ takes into
account more information of the visibility graph structure than the
KLD using degree distributions, providing a closer bound to the physical dissipation
as it is expected by the chain rule \cite{cover}, $D[P_{\rm out}(k,k')||P_{\rm in}(k,k')]\geq D[P_{\rm
out}(k)||P_{\rm in}(k)]$. The improvement is significant in some
situations. Consider for instance the flashing ratchet with a force
opposite to the net current on the system \cite{Parrondo2}. The
current vanishes for a given value of the force usually termed as
\emph{stalling force}. When the force reaches this value, the system
is still out of equilibrium ($V>0$) and it is
 therefore time irreversible, but no current of particles is observed if we describe
 the dynamics of the ratchet with partial
information given by the position $x$.

\begin{figure}[h]
\centering
\includegraphics[width=0.6\textwidth]{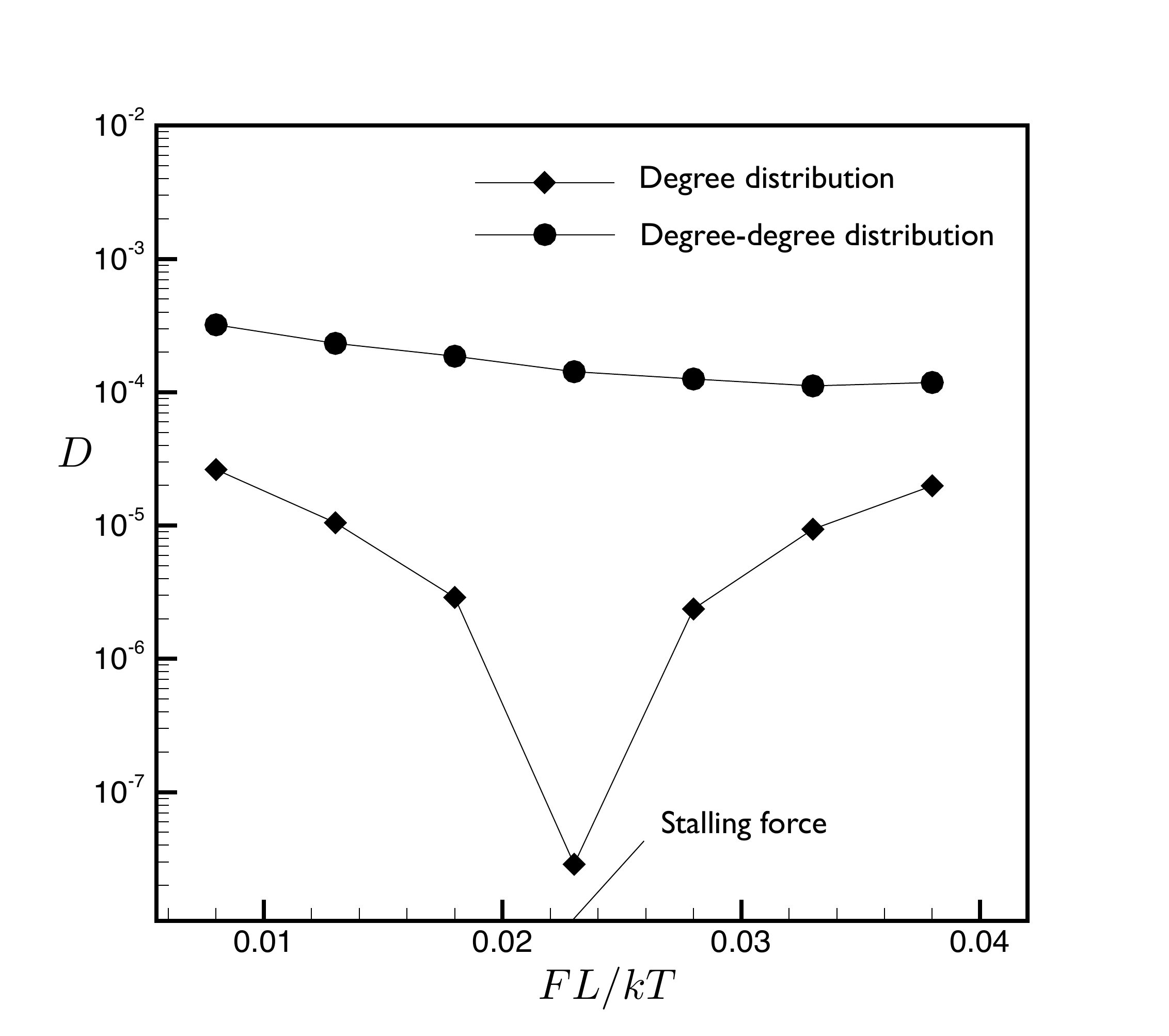}
\caption{Irreversibility measures $D[P_{\rm out}(k)||P_{\rm in}(k)]$
and $D[P_{\rm out}(k,k')||P_{\rm in}(k,k')]$  in the
flashing ratchet ($r=2, V=2kT$) as a function of  $FL/kT$. Here, $F$ is the applied force and $L$ is the spatial
period of the ratchet, which in this case is equal to $1$. For each value of the force, we make use of
a single stationary series of size $N=10^6$ containing
partial information (the state information is removed).} \label{fstall}
\end{figure}

In Fig. \ref{fstall} we show how $D[P_{\rm out}(k)||P_{\rm in}(k)]$
tends to zero when the force approaches to the stalling value.
Therefore, our measure of irreversibility \eqref{dkl} fails in this
case, as do other KLD estimators based on local flows or currents
\cite{Parrondo2}. However, $D[P_{\rm out}(k,k')||P_{\rm in}(k,k')]$
captures the irreversibility of the time series, and yields a
positive value at the stalling force.

\subsection{Chaotic series}

We have applied our method to several chaotic series and found that
it is able to distinguish between dissipative and conservative
chaotic systems. Dissipative chaotic systems are those that do not
preserve the volume of the phase space, and they produce
irreversible time series. This is the case of chaotic maps in which
entropy production via instabilities in the forward time direction
is quantitatively different to the amount of past information lost.
In other words, those whose positive Lyapunov exponents, which
characterize chaos in the forward process, differ in magnitude with
negative ones, which characterize chaos in the backward process
\cite{kennel}. In this section we analyze several chaotic maps and
estimate the degree of reversibility of their associated time series
using our measure, showing that for dissipative chaotic series it is
positive while it vanishes for an example of conservative chaos.

\subsubsection{The Logistic map at $\mu=4$ is irreversible: analytical derivations}
\begin{figure}[h]
\centering
\includegraphics[width=0.7\textwidth]{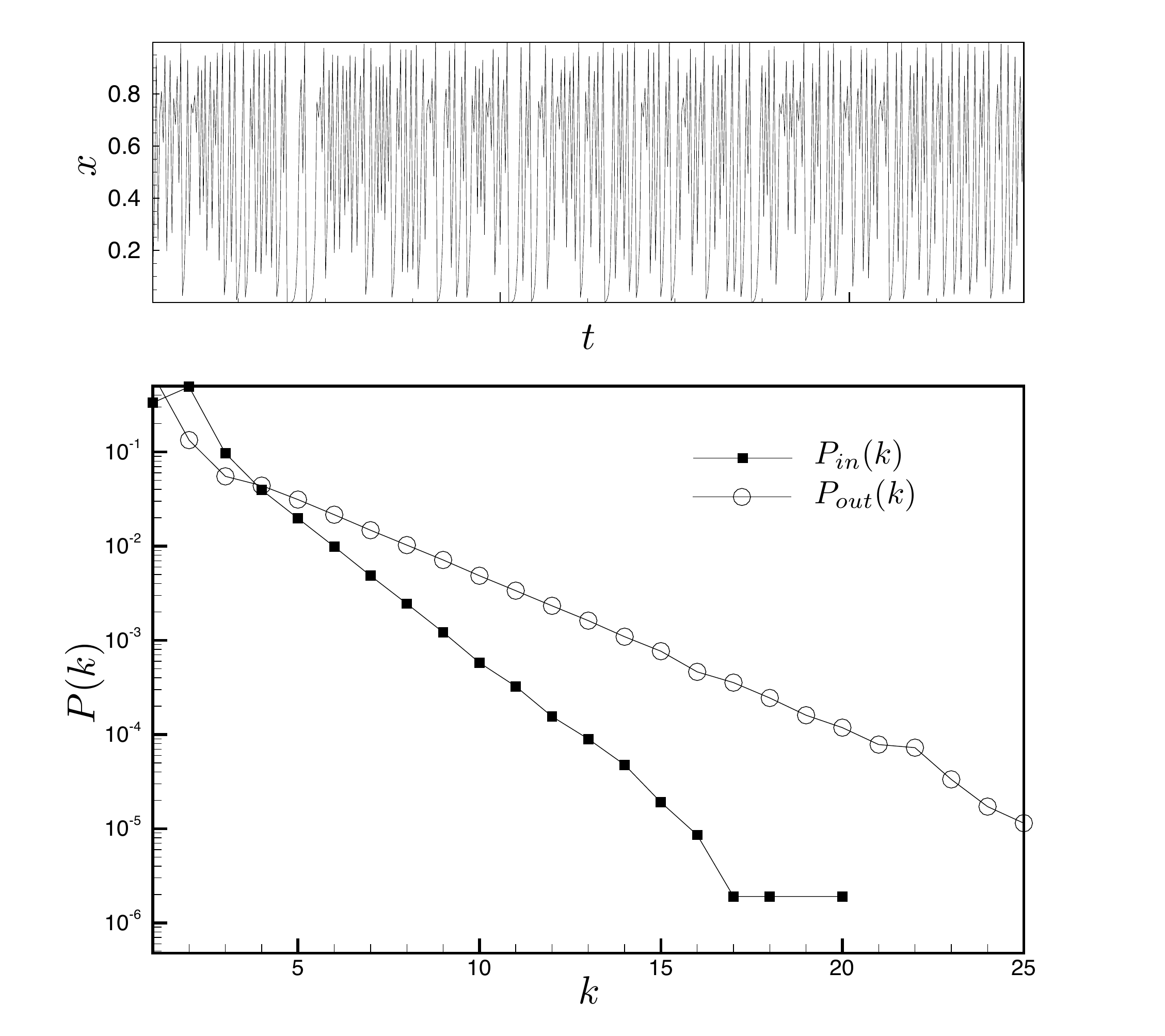}
\caption{\emph{Top}: A sample chaotic time series extracted from a
fully chaotic Logistic map $x_{t+1}=4x_t(1-x_t)$. \emph{Bottom}:
The \emph{in} and \emph{out} degree distributions of the DHVg
associated to the chaotic series. The process is irreversible and
the graph degree distributions are clearly different. The
deviation is measured through the KLD, which is positive in this case (see table \ref{table1}).}
\label{caos}
\end{figure}

\begin{figure}[h]
\centering
\includegraphics[width=0.6\textwidth]{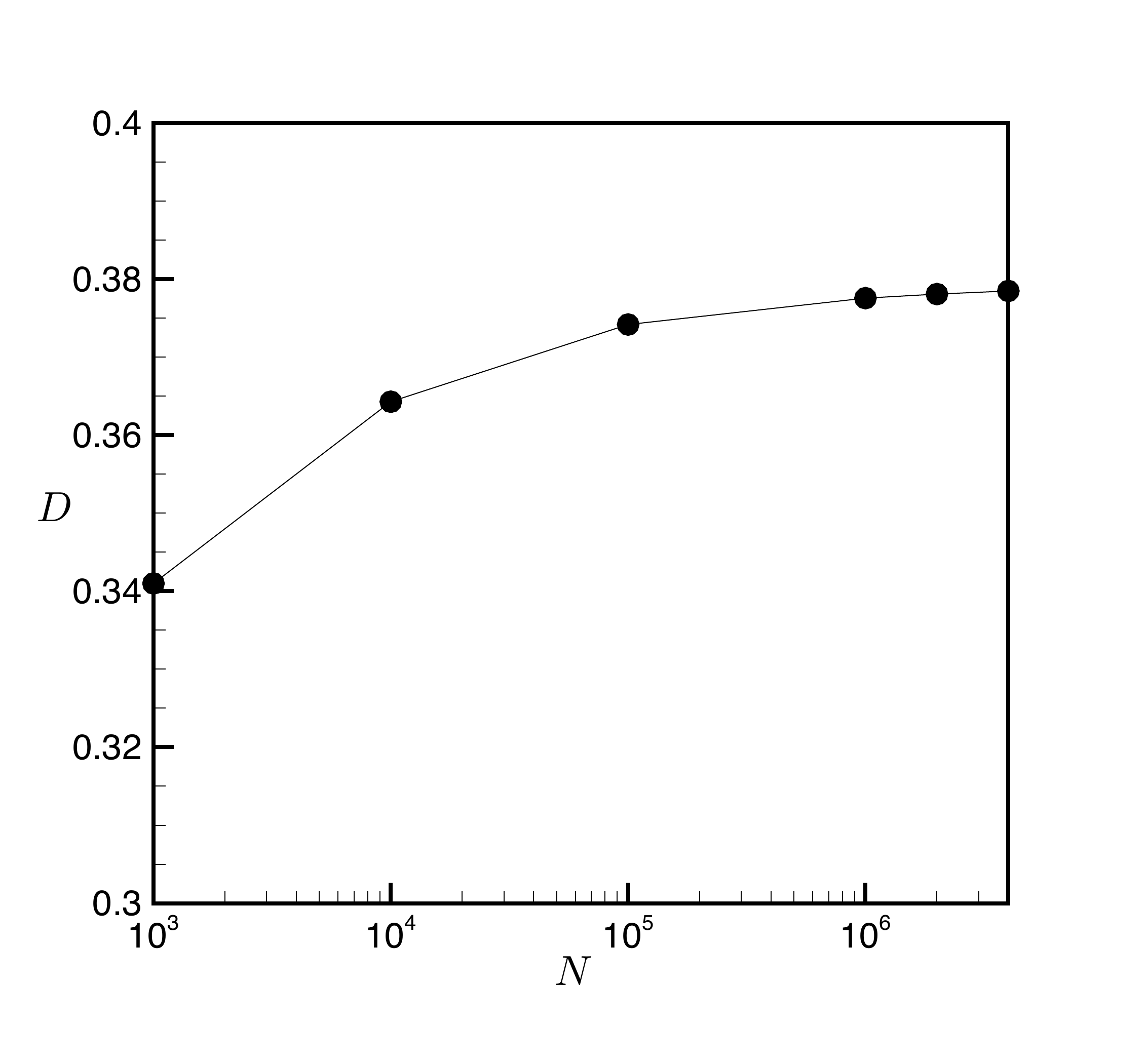}
\caption{Log-log plot of $D[P_{\rm out}(k)||P_{\rm in}(k)]$ of the
graph associated to a fully chaotic Logistic map
$x_{t+1}=4x_t(1-x_t)$, as a function of the series size $N$ (dots
are the result of an ensemble average over different
realizations). Our irreversibility  measure converges
with series size to an asymptotical nonzero value.} \label{finitesize2}
\end{figure}

For illustrative purposes, in figure \ref{caos} we have plotted the
\emph{in} and \emph{out} degree distributions of the DHVg associated
to a paradigmatic dissipative chaotic system: the Logistic map at
$\mu=4$. There is a clear distinction between both distributions, as
it is quantified by the KLD, which in this case is positive both for
degree and degree-degree cases (see table \ref{table1}).
Furthermore, in figure \ref{finitesize2} we make a finite size
analysis in this particular case, showing that our measure quickly
converges to an asymptotic value which clearly deviates from zero,
at odds with reversible processes.

Recall that in section III we proved analytically that for a random
uncorrelated process $D[P_{\rm out}(k)||P_{\rm in}(k)]=0$, since
$P_{\rm in}(k)=P_{\rm out}(k)$. Proving a similar result for a
generic irreversible process is a major challenge, since finding out
exact results for the entire degree distributions is in general
difficult \cite{toral}. However, note that the KLD between two
distributions is zero  if and only if the distributions are the same
in the entire support. Therefore, if we want to prove that this
measure is strictly positive, it is sufficient to find that $P_{\rm
in}(k)\neq P_{\rm out}(k)$ for some value of the degree $k$. Here we
take advantage of this fact to provide a rather general recipe to
prove that a chaotic system is irreversible.

Consider a time series $\{x_t\}_{t=1,...,N}$ with a joint
probability distribution $f(x_1,x_2,...,x_N)$ and support $(a,b)$,
and denote $x_{t-1},x_t, x_{t+1}$ three (ordered) generic data of
the series. By construction,
\begin{eqnarray}
P_{\rm out}(k=1)&=&P(x_{t}\leq x_{t+1})=\int_a^b dx_{t} \int_{x_t}^b dx_{t+1} f(x_{t},x_{t+1}) ,\nonumber\\
P_{\rm in}(k=1)&=& P(x_{t-1}>x_t)=1-P_{\rm out}(k=1).
\end{eqnarray}

The probability that $k_{\rm out}=1$  ($k_{\rm in}=1$) is actually
the probability that the series increases (decreases) in one step.
This probability is independent of time, because we consider
stationary series. If the chaotic map is of the form
$x_{t+1}=F(x_t)$, it is Markovian, and the preceding equations
simplify:
\begin{equation}
P_{\rm out}(k=1) =\int_a^b dx_{t} \int_{x_t}^b dx_{t+1} f(x_{t})f(x_{t+1}|x_t). \label{p1}
\end{equation}
  For chaotic dynamical systems whose trajectories are in
the attractor, there exists an invariant probability measure that
characterizes the long-term fraction of time spent by the system in
the various regions of the attractor. In the case of the Logistic
map
\begin{equation}
F(x_t)=\mu x_t(1-x_t)
\label{log_eq}
\end{equation}
with parameter  $\mu = 4$, the attractor is the whole interval $[0,1]$ and the
probability measure $f(x)$ corresponds to
\begin{equation}
f(x)=\frac{1}{\pi \sqrt{x(1-x)}}. \label{rho}
\end{equation}
Now, for a deterministic system, the transition probability is
simply
\begin{equation}
f(x_{t+1}|x_t)=\delta(x_{t+1}-F(x_t)), \label{rho2}
\end{equation}
where $\delta(x)$ is the Dirac delta distribution. Equations (\ref{p1}) for the Logistic map with $\mu=4$ and
$x \in [0,1]$ reads
\begin{eqnarray}
P_{\rm out}(k=1)=\int_{0}^1 dx_{t}\int_{x_t}^1 dx_{t+1} f(x_{t})\delta(x_{t+1}-F(x_{t})).\label{log2}
\end{eqnarray}
Notice that, using the properties of the Dirac delta distribution,
$\int_{x_t}^1 \delta(x_{t+1}-F(x_t))dx_{t+1}$ is equal to one iff
$F(x_t)\in[x_t,1]$, what happens iff $0<x_t<3/4$, and it is zero
otherwise. Therefore the only effect of this integral is to
restrict the integration range  of $x_t$ to be $[0,3/4]$. Equation
(\ref{log2}) reduces to

\begin{equation}
P_{\rm out}(k=1)= \int_0^{3/4}dx_t f(x_t)=2/3, \label{poutlog}
\end{equation}
and $P_{\rm in}(k=1)=1/3$.
 We conclude that $P_{\rm out}(k)\neq P_{\rm in}(k)$ for the
Logistic map and hence the KLD measure based on degree distributions
is positive. Recall that $P_{\rm out}(k=1)= 2/3$ is the probability that the series exhibits a positive jump ($x_{t}>x_{t-1}$) once in the attractor. These positive jumps must be smaller in size than the negative jumps because, once in the attractor, $\langle x_{t}\rangle$ is constant. The irreversibility captured by the difference between $P_{\rm out}(k=1)$ and $P_{\rm in}(k=1)$ is then the asymmetry of the probability distribution of the slope $x_{t}-x_{t-1}$ of the original time series. The KLD of the degree distributions given by \eqref{dkl} clearly goes beyond this simple signature of irreversibility and can capture more complex and long-range traits.
\begin{table}
\begin{ruledtabular}
\begin{tabular}{cccccddd}
\textbf{Series description}&$D[P_{\rm out}(k)||P_{\rm in}(k)]$&$D[P_{\rm out}(k,k')||P_{\rm in}(k,k')]$\\
\hline
\emph{Reversible Stochastic Processes}\\
\\
$U[0,1]$ uncorrelated&$3.88\cdot10^{-6}$&$2.85\cdot 10^{-4}$\\
Ornstein-Uhlenbeck ($\tau=1.0$)&$7.82\cdot10^{-6}$&$1.52\cdot 10^{-4}$\\
Long-range correlated \\stationary process ($\gamma=2.0$)&$1.28\cdot10^{-5}$&$2.0\cdot 10^{-4}$\\
\hline
\emph{Dissipative Chaos}\\
\\
Logistic map ($\mu=4$)&0.377&2.978\\
$\alpha$ map ($\alpha=3$)&0.455&3.005\\
$\alpha$ map ($\alpha=4$)&0.522&3.518\\
Henon map ($a=1.4, b=0.3$)&0.178&1.707\\
Lozi map &0.114&1.265\\
Kaplan Yorke map &0.164&0.390\\
\hline
\emph{Conservative Chaos}\\
\\
Arnold Cat map &$1.77\cdot10^{-5}$&$4.05\cdot 10^{-4}$\\
\end{tabular}
\end{ruledtabular}
\caption{\label{table1} Values of the irreversibility measure
associated to the degree distribution $D[P_{\rm out}(k)||P_{\rm
in}(k)]$ and the degree-degree distribution $D[P_{\rm
out}(k,k')||P_{\rm in}(k,k')]$ respectively, for the visibility
graphs associated to series of $10^6$ data generated from
reversible and irreversible processes. In every case chain rule is
satisfied, since $D[P_{\rm out}(k,k')||P_{\rm in}(k,k')]\geq
D[P_{\rm out}(k)||P_{\rm in}(k)]$. Note that that the method
correctly distinguishes between reversible and irreversible
processes, as KLD vanishes for the former and it is positive for
the latter.}
\end{table}
\subsubsection{Other chaotic maps}
For completeness, we consider  other   examples of dissipative
chaotic systems analyzed in \cite{sprott2}:

\begin{enumerate}
\item \emph{The $\alpha$-map}: $x_{t+1}=1-|2x_t-1|^{\alpha}$,
which reduces to the Logistic and tent maps in their fully
chaotic region for $\alpha=2$ and $\alpha=1$ respectively. We
analyze this map for $\alpha=3,4$.

\item \emph{The 2D H\'{e}non map}: $x_{t+1}=1+y_t-ax_t^2$, $y_{t+1}=bx_t$,
in the fully chaotic region ($a=1.4$, $b=0.3$).

\item \emph{ The Lozi map}: a piecewise-linear variant
of the H\'{e}non map given by $x_{t+1}=1+y_n-a|x_t|,\
y_{t+1}=bx_t$ in the chaotic regime ($a=1.7$ and $b=0.5$).

\item \emph{The Kaplan-Yorke map}: $x_{t+1}=2x_t \; \textrm{mod}(1),
 y_{t+1}=\lambda y_t + \cos(4\pi x_t) \; \textrm{mod}(1)$.

 \end{enumerate}

 We generate stationary time series with these maps and take data
once the system is in the corresponding attractor. In table
\ref{table1} we show the value of the KLD between the in and out
degree and degree-degree distributions for these series. In every
case, we find an asymptotic positive value, in agreement with the
conjecture that dissipative chaos is indeed time irreversible.

\begin{figure}[h]
\centering
\includegraphics[width=0.6\textwidth]{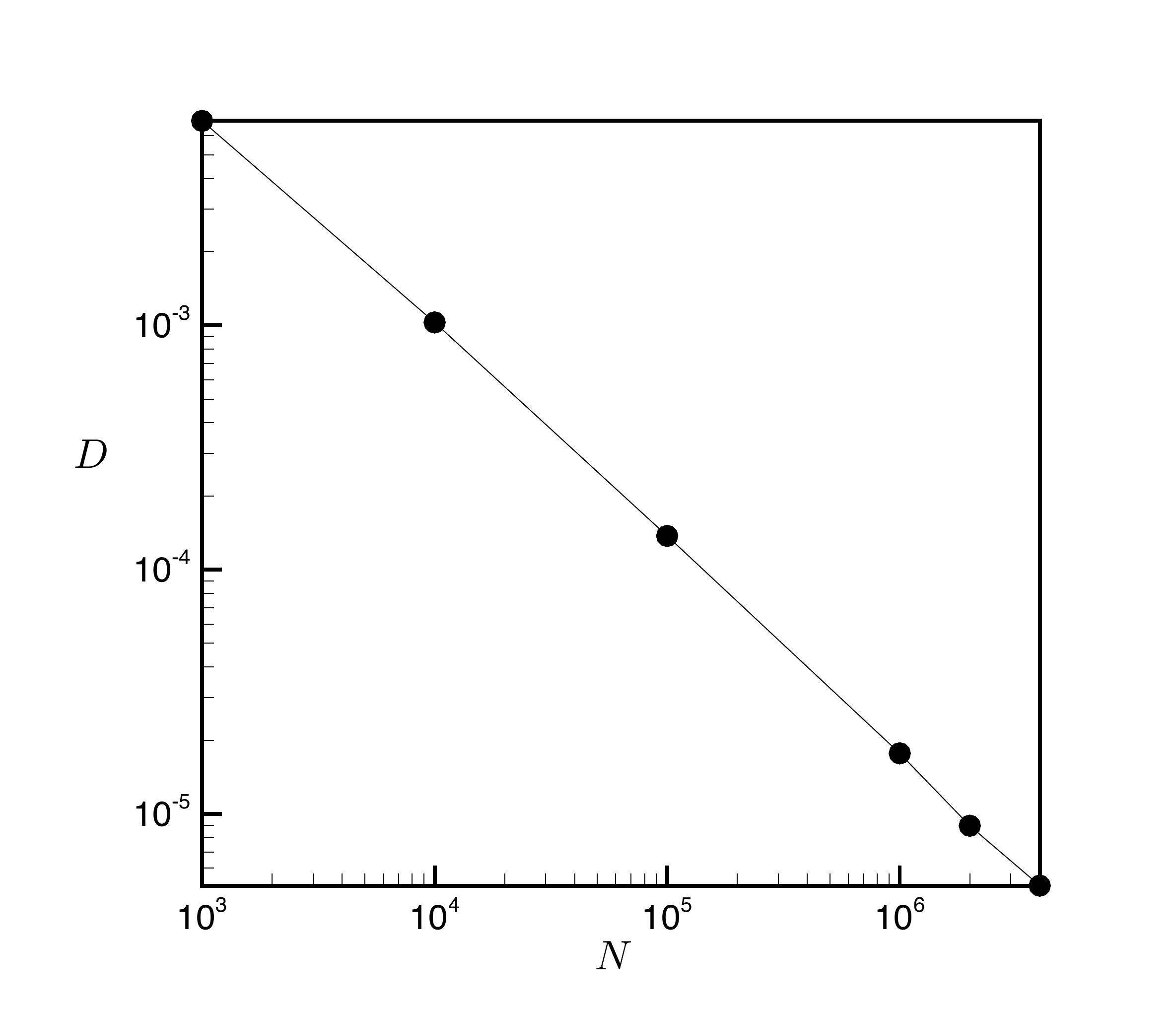}
\caption{Log-log plot of $D[P_{\rm out}(k)||P{in}(k)]$ of the graph associated to the Arnold cat map
 as a function of the series size $N$ (dots are the result of an ensemble average over different realizations).
 Note that the irreversibility measure decreases with series size,
and asymptotically tends to zero, which suggests that this chaotic map is reversible.} \label{finitesize3}
\end{figure}

Finally, we also consider the \emph{Arnold cat map}:
$x_{t+1}=x_t+y_t \ \textrm{mod}(1), y_{t+1}=x_t+2y_t \
\textrm{mod}(1)$. At odds with previous dissipative maps, this is an
example of a \emph{conservative} (measure-preserving) chaotic system
with integer Kaplan-Yorke dimension \cite{sprott2}. The map has two
Lyapunov exponents which coincide in magnitude
$\lambda_1=\ln(3+\sqrt5)/2=0.9624$ and
$\lambda_2=\ln(3-\sqrt5)/2=-0.9624$. This implies that the amount of
information created in the forward process ($\lambda_1$) is equal to
the amount of information created in the backwards process
($-\lambda_2$), therefore the process is time reversible. In figure
\ref{finitesize3} we show that $D[P_{\rm out}(k)||P_{\rm in}(k)]$
for a time series of this map
 asymptotically tends to zero with series size, and the same happens
 with the degree-degree distributions (see table \ref{table1}).
This correctly suggests that albeit chaotic, the map is statistically time reversible.

\subsection{Irreversible chaotic series polluted with noise}
\begin{figure}[h]
\centering
\includegraphics[width=0.6\textwidth]{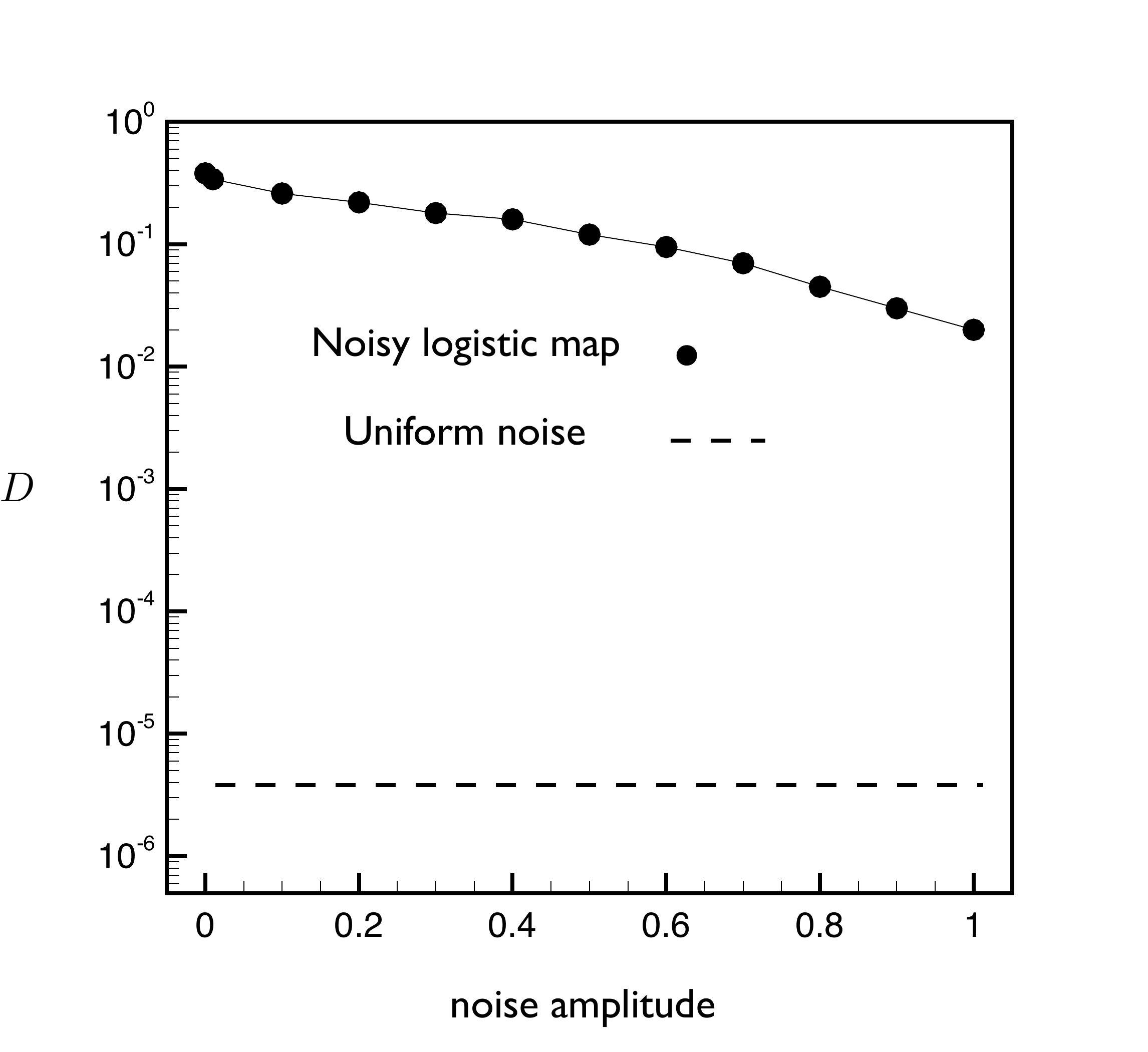}
\caption{Semi-log plot of $D[P_{\rm out}(k)||P_{\rm in}(k)]$ of the
graph associated to series of $10^6$ data extracted from a fully
chaotic Logistic map $x_{t+1}=4x_t(1-x_t)$ polluted with extrinsic
white uniform noise $U[-0.5,0.5]$, as a function of the noise
amplitude. The corresponding KLD value of a uniform
series is plotted for comparison, which is five orders of
magnitude smaller even when the chaotic signal is polluted with an
amount of noise of the same amplitude. This suggests that our measure is robust against noise.} \label{robust}
\end{figure}
Standard time series analysis methods  evidence problems when
noise is present in chaotic series. Even a small amount of noise
can destroy the fractal structure of a chaotic attractor and
mislead the calculation of  chaos indicators such as the
correlation dimension or the Lyapunov exponents \cite{noise}.
 In order to check if our method is robust, we add an amount of
 white noise (measurement noise) to a signal extracted from a fully
chaotic Logistic map ($\mu=4.0$). In figure \ref{robust} we plot
$D[P_{\rm out}(k)||P_{\rm in}(k)]$  of its associated visibility
graph as a function of the noise amplitude (the value corresponding
to a pure random signal is also plotted for comparison). The KLD of
the signal polluted with noise is significantly greater than zero,
as it exceeds the one associated to the noise in four orders of
magnitude, even when the noise reaches the $100\%$ of the signal
amplitude. Therefore our method correctly predicts that the signal
is irreversible even when adding noise.

\section{Discussion}
In this paper we have introduced a new method to measure time
irreversibility of real valued stationary stochastic time series.
The algorithm proceeds by mapping the series into an alternative
representation, the directed horizontal visibility graph. We have
shown that the Kullback-Leibler divergence (KLD) between the $in$
and $out$ degree distributions calculated on this graph is a
measure of the irreversibility of the series.

We have shown that the difference between the $in$ and $out$
distributions at $k=1$ measures the asymmetry in the PDF of the slope of the series.
The degree, however, contains information of
long-range correlations and structures in the series. In particular, the degree-degree distribution can detect the irreversibility in a series with a symmetric slope, as we have shown for the flashing ratchet at stall force.

Our technique  discriminates between conservative and dissipative chaotic maps. The
method has been validated by studying both reversible (uncorrelated
and linearly correlated stochastic processes as well as conservative
chaotic maps) and irreversible (out-of-equilibrium physical
processes and dissipative chaotic maps) series.

We have also shown that the method is robust against noise, in the
sense that irreversible signals are well characterized even when
these signals are polluted with a significant amount of
(reversible) noise. It is also worth emphasizing that it lacks a
symbolization process, and hence it can be applied directly to any
kind of real-valued time series. This makes our technique of
potential interest for several communities. This includes for
instance biological sciences, where there is not such a simple
tool to discriminate between time series generated by active
(irreversible) and passive (reversible) processes.

\textbf{Acknowledgments} We acknowledge financial support from
grants MODELICO, Comunidad de Madrid; FIS2009-13690 (LL, AN and
BL) and MOSIACO (ER and JMRP), Ministerio de Educaci\'on.


\end{document}